\documentclass[11pt,a4paper]{article}
\usepackage{jheppub}
\usepackage{axodraw}
\usepackage{base}
\usepackage{fmslash}

 \usepackage{ulem}% fuer \sout{} um Sachen durchzustreichen. Spaeter wieder auskommentieren, da auch \emph{} Sachen im
\newcommand{\Lcal}{\mathcal{L}}
\DeclareMathOperator{\br}{Br}


\preprint{TTP13-014}

\title{Status of the two-Higgs-doublet model of type II} 

\author[a]{Otto Eberhardt,}
\author[a]{Ulrich  Nierste,}
\author[a]{and Martin Wiebusch}

\affiliation[a]{Institut f\"ur Theoretische Teilchenphysik, Karlsruhe
  Institute of Technology, Engesserstra\ss e 7, 76128 Karlsruhe,
  Germany}

\emailAdd{otto.eberhardt@kit.edu}
\emailAdd{ulrich.nierste@kit.edu}
\emailAdd{martin.wiebusch@kit.edu}

\abstract{We determine the allowed parameter space of the $CP$-conserving
  two-Higgs-doublet model (2HDM) of type II with a softly broken $Z_2$ symmetry.
  Our analysis includes theoretical constraints from vacuum stability and
  perturbativity as well as experimental constraints from signal strengths of
  the \unit{126}{GeV} Higgs boson, the non-observation of additional Higgs
  resonances and electroweak precision and flavour observables. If the
  \unit{126}{GeV} resonance is interpreted as the light $CP$-even Higgs boson of
  the 2HDM our analysis shows that scenarios where the couplings of this boson
  deviate substantially from those of the SM Higgs boson are disfavoured at one
  standard deviation and completely excluded for small values of $\tan\beta$. We
  also discuss bounds on the masses of the heavy 2HDM Higgs bosons and their
  implications for the possible decay modes of these particles. We find that
  the region in which both non-standard neutral Higgs bosons are simultaneously
  lighter than \unit{300}{GeV} is excluded at two standard deviations.}

\begin{document}

\maketitle

\section{Introduction}
The LHC experiments ATLAS and CMS have discovered a neutral boson whose
properties comply with those of the Standard-Model (SM) Higgs boson
\cite{Chatrchyan:2012ufa, Aad:2012tfa}.  Moreover, the data on the Higgs signal
strength have permitted to exclude a sequential fourth fermion generation at the
level of 5 standard deviations \cite{Djouadi:2012ae, Kuflik:2012ai,
  Eberhardt:2012sb, Buchkremer:2012yy, Eberhardt:2012ck,
  Eberhardt:2012gv}. Similarly to the number of fermion generations, the
structure of the Higgs sector is an ad-hoc feature of the SM: While a single
Higgs doublet is sufficient to break the electroweak symmetry, there are no
fundamental reasons forbidding a richer Higgs sector. From a purely
phenomenological point of view, the logical next step after the discovery of a
Higgs boson is to address the question whether it really is ``the'' Higgs
boson. If nature has opted for an extended Higgs sector, the latter will
influence precision observables through loops with extra Higgs bosons. In order
to assess the viable parameter space of a given extension of the SM, one must
perform a global fit in the extended model which includes all relevant
theoretical and experimental constraints.  In this paper we perform such an
analysis for the popular two-Higgs-doublet model (2HDM) \cite{Lee:1973iz} of
type II \cite{Glashow:1976nt, Donoghue:1978cj}, in the widely-studied version
without $CP$ violation in the Higgs potential.

The presence of an additional Higgs doublet implies the existence of three
neutral ($h$, $H$, $A$) and two charged ($H^\pm$) Higgs bosons. The 2HDM of
type-II is designed to avoid flavour-changing couplings of the neutral Higgs
boson by coupling one Higgs doublet solely to up-type and the other one to
down-type fermions. Theoretical constraints on this model come from
the following requirements:
\begin{itemize}
\item the Higgs potential must be bounded from below,
\item neglecting the possibility of a meta-stable vacuum, the minimum of the
  Higgs potential with a vacuum expectation value (VEV) $v=\unit{246}{GeV}$ must
  be the global minimum,
\item the Higgs and Yukawa couplings must be perturbative.
\end{itemize}
The relevant experimental constraints are:
\begin{itemize}
\item the mass and signal strengths of the observed Higgs resonance at
  \unit{126}{GeV},
\item the non-observation of additional Higgs resonances at LEP, Tevatron and
  the LHC,
\item the electroweak precision observables measured at LEP,
\item flavour observables from radiative $B$ decays and $B-\bar B$ mixing.
\end{itemize}
A comprehensive and thorough analysis of constraints from flavour physics has
been performed by the CKMfitter group in \cite{Deschamps:2009rh}.  In our study
we only include the two most relevant flavour observables, namely the branching
ratio of $\bar B\to X_s\gamma$ and the $B_s$-$\bar B_s$ mixing frequency.  After
the Higgs discovery the compatibility of the type-II 2HDM with the observed
Higgs signal strengths and other experimental data has been studied in several
papers \cite{Ferreira:2011aa, Blum:2012kn, Basso:2012st, Cheon:2012rh,
  Carmi:2012in, Ferreira:2012nv, Drozd:2012vf, Chang:2012zf, Chen:2013kt,
  Celis:2013rcs, Giardino:2013bma, Grinstein:2013npa, Shu:2013uua,
  Barroso:2013zxa, Coleppa:2013dya}. However, to our knowledge the analysis
presented here is the first global fit which consistently includes all the
above-mentioned constraints. The Higgs signal strengths provide strong bounds on
the 2HDM parameters which determine the couplings of the light $CP$-even Higgs
$h$, namely on the ratio $\tan\beta$ of the Higgs VEVs and the mixing angle
$\alpha$ of the neutral $CP$-even Higgs bosons.  In this respect, our analysis
updates (some of the) previous studies by using the Higgs data presented at the
Moriond 2013 conference. Furthermore, the above-mentioned theoretical and
experimental constraints allow us to rule out certain combinations of the heavy
2HDM Higgs masses. We also discuss the implications of these limits for the
possible decay modes of heavy 2HDM Higgs bosons. Where appropriate, we compare
our results with those of \cite{Ferreira:2011aa, Blum:2012kn, Basso:2012st,
  Cheon:2012rh, Carmi:2012in, Ferreira:2012nv, Drozd:2012vf, Chang:2012zf,
  Chen:2013kt, Celis:2013rcs, Giardino:2013bma, Grinstein:2013npa, Shu:2013uua,
  Barroso:2013zxa, Coleppa:2013dya}.

Our paper is organised as follows: in Sec.~\ref{sec:model} we provide a brief
overview over the type-II 2HDM and its parametrisation.  In Sec.~\ref{sec:inp}
we discuss the theoretical and experimental constraints included in our analysis
in detail. The results of the global fit are shown in Sec.~\ref{sec:res}.

%%%%%%%%%%%%%%%%%%%%%%%%%%%%%%%%%%%%%%%%%%%%%%%%%%%%%%%%%%%%%%%%%%%%%%%%%%%%%%%%
\section{The Model\label{sec:model}}
%%%%%%%%%%%%%%%%%%%%%%%%%%%%%%%%%%%%%%%%%%%%%%%%%%%%%%%%%%%%%%%%%%%%%%%%%%%%%%%%

The model we consider in this paper is the $CP$-conserving two Higgs doublet
model of type II with a softly broken $Z_2$ symmetry. All relevant details about
this model can be found in \cite{Gunion:2002zf}, whose notational conventions we
follow exactly. The Higgs potential we consider is
\begin{align}
  V
  &=\phantom{{}+{}}
      m_{11}^2\Phi_1^\dagger\Phi_1^{\phantom{\dagger}}
    + m_{22}^2\Phi_2^\dagger\Phi_2^{\phantom{\dagger}}
    - m_{12}^2 (  \Phi_1^\dagger\Phi_2^{\phantom{\dagger}}
                + \Phi_2^\dagger\Phi_1^{\phantom{\dagger}})
    + \tfrac12 \lambda_1(\Phi_1^\dagger\Phi_1^{\phantom{\dagger}})^2
    + \tfrac12 \lambda_2(\Phi_2^\dagger\Phi_2^{\phantom{\dagger}})^2
  \nonumber \\
  &\phantom{{}={}}
    + \lambda_3(\Phi_1^\dagger\Phi_1^{\phantom{\dagger}})
               (\Phi_2^\dagger\Phi_2^{\phantom{\dagger}})
    + \lambda_4(\Phi_1^\dagger\Phi_2^{\phantom{\dagger}})
               (\Phi_2^\dagger\Phi_1^{\phantom{\dagger}})
    + \tfrac12 \lambda_5[  (\Phi_1^\dagger\Phi_2^{\phantom{\dagger}})^2
                         + (\Phi_2^\dagger\Phi_1^{\phantom{\dagger}})^2]
  \eqsep,\label{eq:pot}
\end{align}
where $\Phi_1,\Phi_2$ are the two scalar $SU(2)$ doublets. Under the $Z_2$
symmetry they transform as $(\Phi_1,\Phi_2)\to(-\Phi_1,\Phi_2)$ and the term
with $m_{12}^2$ breaks that symmetry softly. In this paper we only study the
case of unbroken $CP$ symmetry (in the Higgs sector), where we can assume
without loss of generality that
$m_{11}^2,m_{22}^2,m_{12}^2,\lambda_1,\ldots,\lambda_5\in\R$.  At the global
minimum of the potential $V$ the neutral components of $\Phi_1$ and $\Phi_2$
acquire vacuum expectation values (VEVs) $v_1/\sqrt2$ and $v_2/\sqrt2$,
respectively, which are determined by the parameters of the Higgs potential and
must satisfy $v_1^2+v_2^2=(\unit{246}{GeV})^2\equiv v^2$. After trading
$m_{11}$ and $m_{22}$ for $v_1$ and $v_2$ the independent real parameters of the
model are
\begin{equation}\label{eq:params}
  \tan\beta=v_2/v_1
  \eqsep,\eqsep
  m_{12}^2
  \eqsep,\eqsep
  \lambda_1
  \eqsep,\eqsep
  \lambda_2
  \eqsep,\eqsep
  \lambda_3
  \eqsep,\eqsep
  \lambda_4
  \eqsep,\eqsep
  \lambda_5
\end{equation}
and we may assume $\tan\beta>0$ without loss of generality. The physical scalar
spectrum of this model consists of two neutral $CP$-even bosons
$h$ and $H$ (with masses $m_h\leq m_H$), a neutral $CP$-odd boson $A$,
a charged boson $H^+$ and its anti-particle $H^-$. Expressions for the
physical (tree-level) masses of the Higgs bosons in terms of the parameters
\eqref{eq:params} can be found in \cite{Gunion:2002zf}.

The Yukawa Lagrangian of the type-II model is
\begin{equation}
    \Lcal_\text{Yuk}
  = - Y^d_{ij}\bar Q_{Li} \Phi_1 d_{Rj}
    - Y^u_{ij}\bar Q_{Li} \tilde\Phi_2 u_{Rj}
    - Y^l_{ij}\bar L_{Li} \Phi_1 e_{Rj}
    + \text{h.c.}
  \eqsep,
\end{equation}
where $Q_L$ and $L_L$ are the left-handed quark and lepton doublets, $d_R$,
$u_R$ and $e_R$ are the right-handed up-type quark, down-type quark and lepton
singlets, respectively, $Y^u$, $Y^d$ and $Y^l$ are the corresponding Yukawa
coupling matrices, $i,j=1,2,3$ are generation indices and $\tilde\Phi_2\equiv
i\sigma_2\Phi_2^*$ (where $\sigma_2$ is the second Pauli matrix).

The tree-level couplings of the neutral $CP$-even Higgs bosons $h$, $H$ to gauge
bosons and fermions have the same structure as the corresponding couplings of
the SM Higgs boson. The pseudo-scalar $A$ only couples to fermions and the
Feynman rule contains an additional factor $i\gamma_5$. The ratios of coupling
constants (2HDM coupling divided by corresponding SM coupling) only depend on
$\beta$ and the mixing angle $\alpha$ of the neutral $CP$-even 2HDM Higgs
bosons. These ratios are summarised in Table~\ref{tab:couplings}.
\begin{table}
  \centering
  \begin{tabular}{LCCC}
    \hline\hline
      & WW,ZZ & \text{up-type quarks} & \text{down-type quarks, leptons} \\
    \hline
    h & \sin(\beta-\alpha) & \cos\alpha/\sin\beta & -\sin\alpha/\cos\beta \\
    H & \cos(\beta-\alpha) & \sin\alpha/\sin\beta &  \cos\alpha/\cos\beta \\
    A &                  0 &            \cot\beta &             \tan\beta \\
    \hline\hline
  \end{tabular}
  \caption{\label{tab:couplings}
    Tree-level couplings of the neutral 2HDM Higgs bosons to gauge bosons
    and fermions. Each coupling is normalised to the corresponding coupling of
    the SM Higgs boson.}
\end{table}
The relation between $\alpha$ and the parameters \eqref{eq:params} is given in
\cite{Gunion:2002zf}. For the discussion in this paper it is important to note
that the couplings of the light $CP$-even Higgs $h$ (first line of
Table~\ref{tab:couplings}) approach the corresponding SM values for
$\beta-\alpha\to\pi/2$, irrespective of the value of $\beta$.

%%%%%%%%%%%%%%%%%%%%%%%%%%%%%%%%%%%%%%%%%%%%%%%%%%%%%%%%%%%%%%%%%%%%%%%%%%%%%%%%
\section{Theoretical Constraints and Experimental Inputs\label{sec:inp}}
%%%%%%%%%%%%%%%%%%%%%%%%%%%%%%%%%%%%%%%%%%%%%%%%%%%%%%%%%%%%%%%%%%%%%%%%%%%%%%%%

The parameters \eqref{eq:params} are subject to a number of theoretical
constraints. First of all, the potential \eqref{eq:pot} must be bounded from
below. As explained in \cite{Gunion:2002zf}, this is the case if and only if the
following inequalities are satisfied:
\begin{equation}
  \lambda_1>0
  \eqsep,\eqsep
  \lambda_2>0
  \eqsep,\eqsep
  \lambda_3>-\sqrt{\lambda_1\lambda_2}
  \eqsep,\eqsep
  |\lambda_5| < \lambda_3 + \lambda_4 + \sqrt{\lambda_1\lambda_2}
  \eqsep.
\end{equation}
Furthermore, to obtain a stable vacuum state we require that the minimum of the
potential with $v_1^2+v_2^2=(\unit{246}{GeV})^2$ is the global
minimum.\footnote{In doing this we neglect the possibility that our vacuum is
metastable with a lifetime larger than the age of the universe.} As pointed
out recently in \cite{Barroso:2013awa} this requirement leads to the additional
constraint
\begin{equation}
  m_{12}^2\bigl(m_{11}^2-m_{22}^2\sqrt{\lambda_1/\lambda_2}\bigr)
  \bigl(\tan\beta-(\lambda_1/\lambda_2)^{1/4}\bigr) > 0
  \eqsep.
\end{equation}
Finally, if we want to be able to trust perturbative calculations, the magnitude
of the Higgs self-couplings $\lambda_i$ should not be too large. The only
correct way to implement this bound is to compute many higher-order corrections
and assess the convergence of the perturbative series. Here we take the simple
approach of requiring $|\lambda_i|<\lambda_\text{max}$ for $i=1,\ldots,5$ and
some $\lambda_\text{max}>0$. The most conservative choice for
$\lambda_\text{max}$ is $4\pi$, which forces the product of two $\lambda$s and
the loop factor to be smaller than $1$. A study of higher-order corrections for
the case of the SM Higgs sector points to a smaller perturbativity limit, closer
to $\lambda_\text{max}=2\pi$ \cite{Nierste:1995zx}.  To estimate the dependence
of our results on the ultimately arbitrary upper limit $\lambda_\text{max}$ we
show results for $\lambda_\text{max}=2\pi$ and $\lambda_\text{max}=4\pi$.

In addition to these theoretical constraints, we confront the 2HDM described in
Sec.~\ref{sec:model} with the following experimental data:
\begin{itemize}
\item the mass of the observed Higgs resonance
  \begin{equation}
    m_h = \unit{125.96^{+0.18}_{-0.19}}{GeV}
      \eqsep.
  \end{equation}
  This input is a combination of the results presented in
  \cite{ATLAS-CONF-2013-012, ATLAS-CONF-2013-013, CMS-PAS-HIG-13-001,
    CMS-PAS-HIG-13-002}. We always identify the observed Higgs resonance with
  the light $CP$-even 2HDM Higgs boson. Specifically, we neglect the
  possibility that the observed resonance is one of the heavy neutral 2HDM
  Higgs bosons or a degenerate state. See, for instance, \cite{Ferreira:2012my,
    Burdman:2011ki} for a discussion of the former case and
  \cite{Ferreira:2012nv} for the latter.
\item the signal strengths (observed cross section times branching ratio divided
  by SM expectation) of the Higgs resonance at \unit{126}{GeV}.  Our signal
  strength inputs for the different decay modes and, in the case of the
  $\gamma\gamma$ final state, the different event categories defined by the
  experimental groups are summarised in Fig.~\ref{fig:signalstrengths}.  On the
  theory side, the signal strength for a given Higgs production and decay mode
  is given by the product of the corresponding 2HDM/SM ratios of (effective)
  squared couplings. For instance, the gluon fusion contribution
  $\mu(gg\to H\to\gamma\gamma)$ to the $H\to\gamma\gamma$ signal strength is
  given by the product $R_{gg}R_{\gamma\gamma}$, where $R_{gg}$
  ($R_{\gamma\gamma}$) is the square of the effective $Hgg$ ($H\gamma\gamma$)
  coupling calculated in the 2HDM, divided by the same effective coupling
  calculated in the SM.  We use the FeynArts, FormCalc and LoopTools packages
  \cite{Hahn:1998yk, Hahn:2000kx, Hahn:2006qw} to compute $R_{gg}$ and
  $R_{\gamma\gamma}$ at one-loop order. For all other couplings ($HWW$, $HZZ$
  etc.) we use the tree-level values.
  
  To compare quantities such as $\mu(gg\to H\to\gamma\gamma)$ with experimental
  data one needs to know the composition of the Higgs signal in a given final
  state or event category. In other words, one needs to know the fraction with
  which each Higgs production mechanism contributes to the signal seen in each
  final state or event category. In our analysis we use these percentage
  contributions wherever they (or the corresponding selection efficiencies) are
  provided by the experimental groups. Our values for the percentage
  contributions are summarised in Tab.~\ref{tab:pcvalues}. In the case of the
  \unit{8}{TeV} CMS $H\to\tau\tau$ data, we derived the percentage contributions
  from the corresponding selection efficiencies. These efficiencies are
  summarised in Tab.~\ref{tab:tautaueff}.  For the remaining final states we
  assume that the dominant production mode contributes $100\%$ of the signal.
\item limits from searches for heavy neutral Higgs bosons in the $WW$ and $ZZ$
  decay modes. Specifically, we include the (mass dependent) expected limit from
  the CMS $H\to WW\to 2l2\nu$ search (\cite{CMS-PAS-HIG-13-003}, Fig.~9) and the
  expected limit from the CMS $H\to ZZ\to 4l$ search (\cite{CMS-PAS-HIG-13-002},
  Fig.~5, left panel). In the absence of any clear signals for heavy Higgs
  resonances we consider it good practice to use the expected limits instead of
  the observed ones since otherwise the analysis becomes sensitive to background
  fluctuations in the search data. For the same reason we refrain from using the
  signal strength values for heavy Higgs bosons, as provided by the experimental
  groups. 
\item the full set of electroweak pseudo-observables (EWPOs) measured at LEP and
  SLC, as well as the latest results for the $W$ and top mass. We use the same
  inputs as in Table~II of \cite{Eberhardt:2012gv}, and our SM parameters
  ($M_Z$, $m_t$, $\alpha_s$ and $\Delta\alpha^{(5)}_\text{had}(M_Z)$) are fixed
  to the best-fit values from that analysis.  We emphasise that the study of the
  oblique parameters $S$,$T$,$U$ is not sufficient, because the 2HDM involves
  $Z$ vertex corrections \cite{Hollik:1986gg, Hollik:1987fg, Haber:1999zh}. For
  our analysis we have re-calculated the 2HDM contributions to the electroweak
  precision observables at one loop using the FeynArts, FormCalc and LoopTools
  packages \cite{Hahn:1998yk, Hahn:2000kx, Hahn:2006qw}. The results have then
  been combined with the SM contributions (including all available higher-order
  corrections) using the prescription of Ref.~\cite{Gonzalez:2011he}.
  The SM contributions to the EWPOs were calculated with the
  Zfitter software \cite{Bardin:1989tq, Bardin:1999yd, Arbuzov:2005ma},
  with the exception of $R_b$, for which we use the improved results
  from \cite{Freitas:2012sy}. 
  %% Electroweak precision observables in (variants of the) 2HDM
  %% have been calculated in Refs.~\cite{Frere:1982ma, Bertolini:1985ia,
  %% Hollik:1986gg, Hollik:1987fg, Froggatt:1991qw, Haber:1999zh,
  %% LopezVal:2012zb}.  
\item the branching ratio $\br(B\to X_s\gamma)$. We use the theoretical
  calculation of this quantity in the 2HDM in Refs.~\cite{Ciuchini:1997xe,
    Borzumati:1998tg, Borzumati:1998nx, Ciafaloni:1997un, Bobeth:1999ww,
    Hermann:2012fc} and write \cite{Misiak:2006ab}
  \begin{equation}
     \br(\bar B\to X_s\gamma)_{E>E_0}
    =\br(\bar B\to X_ce\bar\nu)_\text{exp}
     \left|\frac{V_{ts}^*V_{tb}}{V_{cb}}\right|^2
     \frac{6\alpha_\text{em}}{\pi\cdot C}[P(E_0)+N(E_0)]
    \eqsep,
  \end{equation}
  where $E_0=\unit{1.6}{GeV}$, $\br(\bar B\to X_ce\bar\nu)_\text{exp}=0.1072$
  (Eq.~183 of \cite{Amhis:2012bh}), $|V_{ts}^*V_{tb}/V_{cb}|^2=0.963$ (text
  before Eq.~1 of \cite{Gambino:2008fj}) and $C=0.546$ (Eq.~7 of
  \cite{Gambino:2008fj}). The dependence on the 2HDM parameters is contained in
  the quantity $[P(E_0)+N(E_0)]$. To evaluate it we use private code provided by
  the authors of \cite{Hermann:2012fc}.  Following the discussion of theoretical
  errors in \cite{Hermann:2012fc} we obtain a statistical error of $3\%$ from
  the uncertainties of the parameters $\br(\bar B\to X_ce\bar\nu)_\text{exp}$,
  $|V_{ts}^*V_{tb}/V_{cb}|^2$ and $C$ and an overall systematic error of $12\%$
  (all other errors from \cite{Gambino:2008fj} added linearly). In our fit, all
  these theoretical errors are reflected by a single multiplicative nuisance
  parameter. Our experimental input for $\br(\bar B\to X_s\gamma)_{E>E_0}$ is
  \cite{Hermann:2012fc}
  \begin{equation}
    \br(\bar B\to X_s\gamma)_{E>E_0}^\text{exp} = (3.37\pm 0.23)\times 10^{-4}
    \eqsep.
  \end{equation}
\item the mass splitting $\Delta m_{B_s}$ in the neutral $B_s$ meson system.
  For the theoretical computation of this quantity we use the expressions
  given in \cite{Abbott:1979dt, Geng:1988bq, Buras:1989ui, Deschamps:2009rh}:
  \begin{equation}
     \Delta m_{B_s}
    =\frac{G_F^2}{24\pi^2}|V_{ts}V_{tb}^*|^2
     \eta_Bm_{B_s}\bar m_t^2f_{B_s}^2\hat B_{B_s}
     (S_{WW} + S_{WH} + S_{HH})
    \eqsep,
  \end{equation}
  where $G_F$ is the Fermi constant and the dependence on the 2HDM parameters is
  in the quantities $S_{WW}$, $S_{WH}$ and $S_{HH}$ (see \cite{Deschamps:2009rh}
  for their definition). The values of the other pre-factors are
  \begin{align*}
    |V_{ts}V_{tb}^*|^2 &= 0.039986 && \text{\cite{CKMfitter:2012mo}} \\
    \eta_B &= 0.551\pm0.0022\,\text{(syst.)}
      &&\text{\cite{Buras:1990fn, Deschamps:2009rh}} \\
    m_{B_s} &= \unit{5.3663}{GeV} &&\text{\cite{Beringer:1900zz}} \\
    \bar m_t &= \unit{166.6}{GeV}\eqsep(\overline{MS}\ \text{scheme})
      &&\text{\cite{Eberhardt:2012gv}} \\
    f_{B_s} &= \unit{[0.229\pm0.002\,\text{(stat.)}
                           \pm0.006\,\text{(syst.)}]}{GeV}
      && \text{\cite{Lenz:2012az}} \\
    \hat B_{B_s} &= 1.322\pm0.026\,\text{(stat.)}\pm0.035\,\text{(syst.)}
       &&\text{\cite{Gamiz:2009ku,CKMfitter:2012mo}}
  \end{align*}
  By adding the statistical errors in quadrature and the systematic errors
  linearly we obtain a relative statistical uncertainty of $2.6\%$ and
  a relative systematic uncertainty of ${}^{+8.0}_{-8.5}\%$. In our fit,
  these theoretical uncertainties are represented by a single multiplicative
  nuisance parameter. Our experimental input for $\Delta m_{B_s}$ is
  \cite{Aaij:2013mpa}
  \begin{equation}
      \Delta m_{B_s}
    = \unit{[17.768\pm0.023\,\text{(stat.)}\pm0.006\,\text{(syst.)}]}%
           {\reciprocal{(\pico\second)}}
    \eqsep.
  \end{equation}
\end{itemize}

Let us briefly comment on our selection of flavour observables. In the type-II
model under consideration, flavour-changing neutral current (FCNC) processes are
sensitive to 2HDM effects for small and very large values of $\tan \beta$: if
$\tan\beta <1$, the charged-Higgs coupling to the top quark is
enhanced. Conversely, for $\tan\beta \gtrsim 40$ the couplings of $H$, $A$,
$H^\pm$ to bottom quarks and tau leptons is of order 1, leading to sizable
effects in \mbox{(semi-)tauonic} $B$ decays \cite{Hou:1992sy, Akeroyd:2003zr,
  Grzadkowski:1992qj, Miki:2002nz, Nierste:2008qe, Kamenik:2008tj, Trine:2008qv,
  Tanaka:2010se, Fajfer:2012vx, Sakaki:2012ft, Becirevic:2012jf, Celis:2012dk,
  Tanaka:2012nw} and $\br(B\to \ell^+\ell^-)$ \cite{Logan:2000iv}.  $\br(B\to
X_s \gamma)$ plays a special role, because it provides a powerful lower bound on
$M_{H^+}$ which is essentially independent of $\tan\beta$, unless $\tan\beta<1$
\cite{Ciuchini:1997xe, Borzumati:1998tg, Borzumati:1998nx, Ciafaloni:1997un,
  Bobeth:1999ww, Hermann:2012fc}.  An early combined analysis of several flavour
observables for $\tan\beta \leq 1$ can be found in Ref.~\cite{Buras:1989ui}.  An
exhaustive analysis of several leptonic and semileptonic meson (and $\tau$)
decays, $B$-$\bar B$ mixing, $\br(B\to X_s \gamma)$, and $Z\to b \bar b$ is
presented in Ref.~\cite{Deschamps:2009rh}. In the present paper we are
interested in the low $\tan\beta$ region where the Higgs signal strengths still
allow large deviations of $\alpha$ from the SM-like limit $\beta-\pi/2$.
Therefore the only flavour observables relevant to our fit are $\br(B\to X_s
\gamma)$ and the mass splitting $\Delta m_{B_s}$ in the neutral $B_s$ meson
system. The ratio $\Delta m_B/\Delta m_{B_s}$ assumes the same value as in the
SM. Therefore we do not need to include the weaker constraint from $\Delta m_B$
in our fit. Furthermore, the value of $|V_{ts}V_{tb}|$ governing both $\br(\bar
B\to X_s\gamma)$ and $\Delta m_{B_s}$ is not changed if one passes from the SM
to the 2HDM: $V_{tb}$ is approximately $1$, $V_{ts}$ is obtained from $V_{cb}$
trough CKM unitarity and the extra 2HDM Higgs bosons have no impact on the
determination of $V_{cb}$. The omission of data on (semi-) tauonic $B$ decays
affects the fit only for large values of $\tan\beta$. Furthermore, the 2HDM of
type II does not alleviate the tensions between the SM and the experimental
world averages of $\br(B\to \tau \nu)$ and $\br(B\to D^{(*)}\tau \nu)$, but
rather worsens the agreement with the data. (For an analysis of these decay
modes in a general 2HDM see Ref.~\cite{Crivellin:2012ye}.)

%%%%%%%%%%%%%%%%%%%%%%%%%%%%%%%%%%%%%%%%%%%%%%%%%%%%%%%%%%%%%%%%%%%%%%%%%%%%%%%%
\section{Results\label{sec:res}}
%%%%%%%%%%%%%%%%%%%%%%%%%%%%%%%%%%%%%%%%%%%%%%%%%%%%%%%%%%%%%%%%%%%%%%%%%%%%%%%%

In this section we present the results of a global fit incorporating the
constraints discussed in the last section. All fits were done with the
\textit{my}Fitter framework \cite{Wiebusch:2012en} and cross-checked with an
independent implementation in the CKMfitter software \cite{Hocker:2001xe}.  All
$p$-values (and the corresponding $1\sigma$, $2\sigma$ and $3\sigma$ exclusion
limits) were computed by applying Wilks' theorem. Although this is common
practice for analyses like the one presented here, it is not clear how reliable
these $p$-values are as the presence of theoretical constraints violates the
underlying assumptions of Wilks' theorem (see \cite{Wiebusch:2012en} for a
discussion). For the present paper, we decided to follow standard practice
and postpone further studies of this issue to a future publication.

\begin{figure}
  \centering
  \includegraphics{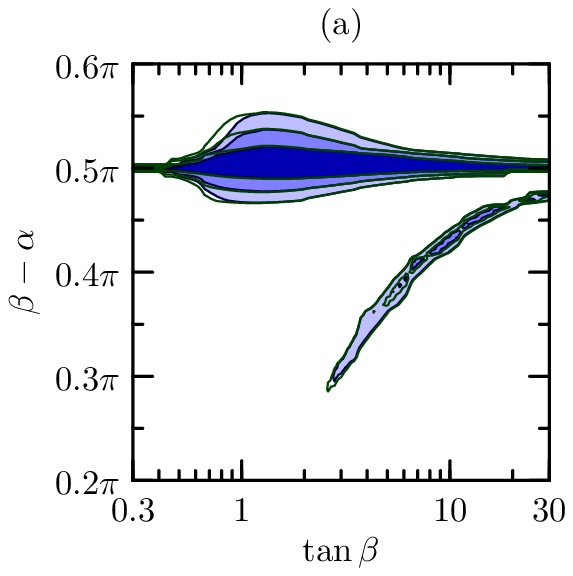}
  \hfil
  \includegraphics{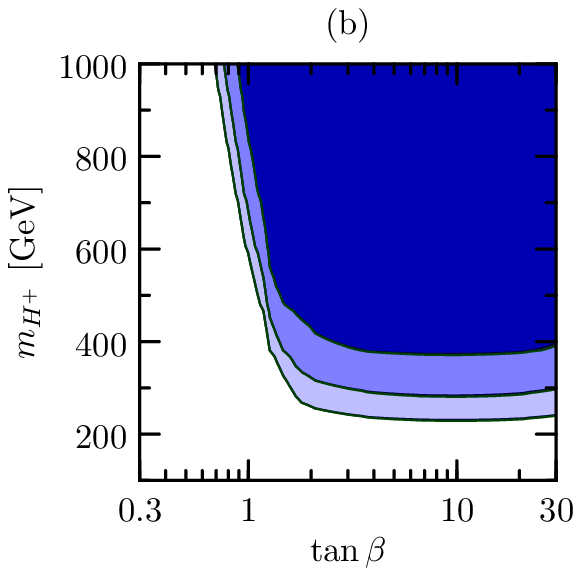}
  \par\smallskip
  \includegraphics{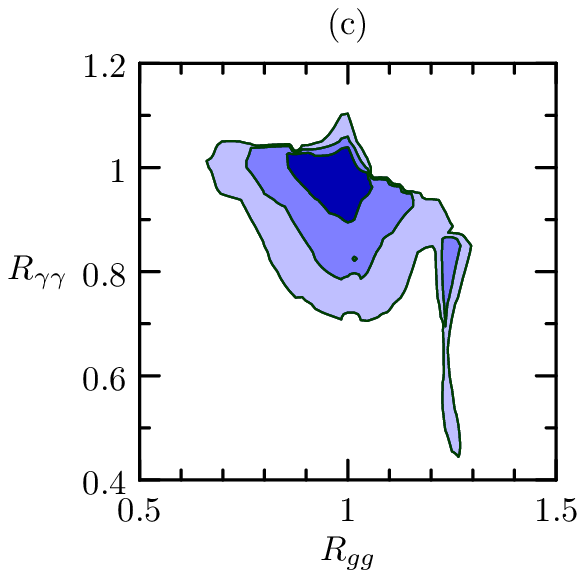}
  \caption{Allowed regions in the $\tan\beta$-$(\beta-\alpha)$ plane (a), the
    $\tan\beta$-$m_{H^\pm}$ plane (b) and the $R_{gg}$-$R_{\gamma\gamma}$ plane
    (c). The shaded blue areas are the regions allowed at one, two and three
    standard deviations (dark to light) for the tight perturbativity constraint
    ($\lambda_\text{max}=2\pi$). The contours of the corresponding regions for
    $\lambda_\text{max}=4\pi$ are indicated by green lines.}
  \label{fig:tb_bma_mHp-full}
\end{figure}

Fig.~\ref{fig:tb_bma_mHp-full}(a) shows the regions in the
$\tan\beta$-$(\beta-\alpha)$ plane allowed at one, two and three standard
deviations. Here and in the following plots, the shaded blue areas show the
results of the fit with the tight perturbativity limit
$\lambda_\text{max}=2\pi$. To gauge the sensitivity of the visible features on
the implementation of the perturbativity bound the contours of the corresponding
areas for the fit with $\lambda_\text{max}=4\pi$ are shown as green lines. The
line with $\beta-\alpha=\pi/2$ corresponds to the case where the couplings of
the light $CP$-even Higgs boson are the same as those of the SM Higgs boson. The
best agreement with the experimental data is found along this line, which just
reflects the fact that all the included experimental data is in good agreement
with the predictions of the SM.  For $\tan\beta<0.6$ the value of $\beta-\alpha$
can not deviate from $\pi/2$ by more than $0.01\pi$. This is a combined effect
of the flavour, EWPO and perturbativity constraints. For small $\tan\beta$ the
observables $\br(\bar B\to X_s\gamma)$, $\Delta m_{B_s}$ and $R_b$ receive large
corrections from charged Higgs diagrams and thus force $m_{H^\pm}$ to large
values. In this limit the perturbativity bounds force $\alpha$ to be close to
$\beta-\pi/2$. For $\tan\beta>5$ there is a thin strip allowed at two standard
deviations, where $\beta-\alpha$ can be as low as $0.4\pi$. The best-fit
scenarios in this strip feature relatively small masses of the charged and
$CP$-even Higgs bosons. For example, we obtain the following best-fit parameters
for $\beta-\alpha$ fixed at $0.4\pi$:
\begin{gather}
  \tan\beta=6.5
  \eqsep,\eqsep
  m_{12}=\unit{185}{GeV}
  \eqsep,\nonumber\\
  m_H=\unit{476}{GeV}
  \eqsep,\eqsep
  m_A=\unit{737}{GeV}
  \eqsep,\eqsep
  m_{H^\pm}=\unit{440}{GeV}
  \eqsep.\label{eq:bestfit}
\end{gather}

Fig.~\ref{fig:tb_bma_mHp-full}(b) shows the allowed regions in the
$\tan\beta$-$m_{H\pm}$ plane.  The exclusions in this plot are essentially due
to the flavour observables $\br(\bar B\to X_s\gamma)$ and $\Delta m_{B_s}$ as
well as the hadronic $Z\to b\bar b$ branching ratio $R_b$.  All these
observables get contributions from charged Higgs diagrams which are proportional
to positive powers of $\cot\beta$. Suppressing these terms for small values of
$\tan\beta$ requires very large values of $m_{H^\pm}$, so that charged Higgs
masses below \unit{1}{TeV} are excluded for $\tan\beta\lesssim0.8$. The
$\beta$-independent terms in $\br(\bar B\to X_s\gamma)$ lead to an absolute
lower limit of \unit{322}{GeV} at two standard deviations and approximately
\unit{400}{GeV} at one standard deviation.  This limit is the main reason for
the fact that the lower strip in Fig.~\ref{fig:tb_bma_mHp-full}(a) is
disfavoured at one standard deviation. If we remove the flavour observables and
$R_b$ from our fit we confirm the results of previous analyses
(e.g.\ \cite{Grinstein:2013npa, Barroso:2013zxa}) where the lower strip in
Fig.~\ref{fig:tb_bma_mHp-full}(a) is still allowed at one standard deviation.
Also note that the green lines in Fig.~\ref{fig:tb_bma_mHp-full}(b) exactly
coincide with the boundaries of the blue regions, which means that the limits
shown in this plot are insensitive to the implementation of the perturbativity
bound. The pattern of Fig.~\ref{fig:tb_bma_mHp-full}(b) is the same as the
one found in \cite{Deschamps:2009rh}, but of course the newer data and the
NNLO result used by us lead to a tighter lower bound on $m_{H^\pm}$.

Limits for the tree-level couplings of $h$ to fermions, $W$ and $Z$ bosons can
easily be extracted from Fig.~\ref{fig:tb_bma_mHp-full}(a) and
Tab.~\ref{tab:couplings}. The relations between the 2HDM parameters and the
one-loop effective $hgg$ and $h\gamma\gamma$ couplings are more complicated.
Fig.~\ref{fig:tb_bma_mHp-full}(c) shows exclusion limits in the
$R_{gg}$-$R_{\gamma\gamma}$ plane, where $R_{gg}$ and $R_{\gamma\gamma}$ are the
(2HDM/SM) ratios of squared effective $hgg$ and $h\gamma\gamma$ couplings,
respectively. We see that the favoured region is centred around
$R_{gg}=R_{\gamma\gamma}=1$, i.e.\ the SM limit. In addition, there is a region
around $R_{gg}=1.25$ and $R_{\gamma\gamma}=0.8$ which is allowed at two standard
deviations. This region directly corresponds to the lower strip in
Fig.~\ref{fig:tb_bma_mHp-full}(a). The enhancement of the $hgg$ coupling is due
to the constructive interference between the top and bottom-loop contribution
and depends only on $\tan\beta$ and $\beta-\alpha$. Using the expressions in
\cite{Ellis:1975ap, Spira:1995rr} for the fermion loop diagram,
$m_h=\unit{126}{GeV}$, $m_t=\unit{174}{GeV}$, $m_b=\unit{4.2}{GeV}$ and coupling
modification factors from Tab.~\ref{tab:couplings} we find
\begin{equation}
  R_{gg}
  \approx  1.107\frac{\cos^2\alpha}{\sin^2\beta}
         + 0.008\frac{\sin^2\alpha}{\cos^2\beta}
         + 0.115\frac{\sin(2\alpha)}{\sin(2\beta)}
  \eqsep.
\end{equation}
For the $\tan\beta$ and $\beta-\alpha$ values from \eqref{eq:bestfit} this gives
$R_{gg}\approx 1.23$. The effective $h\gamma\gamma$ coupling receives
contributions from fermion, $W$ boson and charged Higgs loops. For the
parameters \eqref{eq:bestfit} we obtain $R_{\gamma\gamma}\approx 0.77$.  The
decrease with respect to the SM is due to the fact that the $W$ loop
contribution is multiplied with the factor $\sin(\beta-\alpha)$, which is
approximately $0.95$ for the parameters in \eqref{eq:bestfit}.  The modification
of the $ht\bar t$ coupling and the charged Higgs contribution are negligible at
this parameter point.

\begin{figure}
  \centering
  \includegraphics{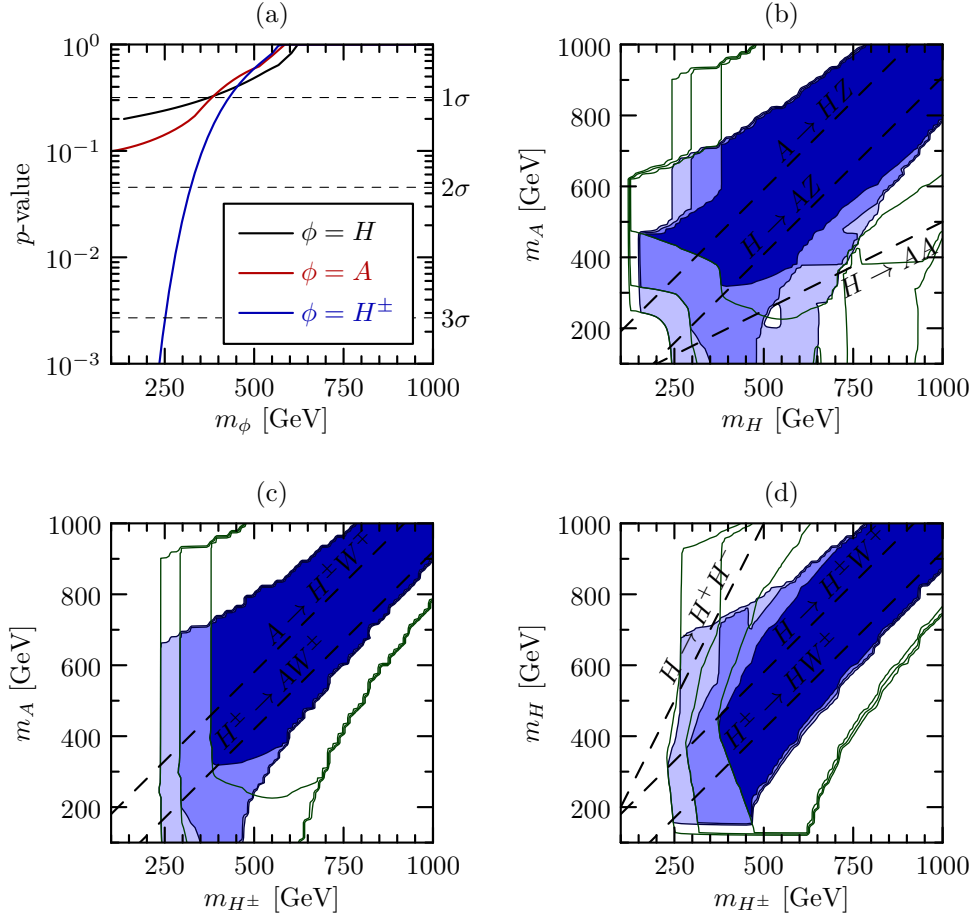}
  \caption{(a) shows $p$-values for the masses of the heavy 2HDM Higgs bosons.
    (b) to (d) shows allowed regions in the $m_H$-$m_A$, $m_{H^\pm}$-$m_A$ and
    $m_{H^\pm}$-$m_H$ planes, respectively. The shaded blue areas are the
    regions allowed at one, two and three standard deviations (dark to light)
    for the tight perturbativity constraint ($\lambda_\text{max}=2\pi$). The
    contours of the corresponding regions for $\lambda_\text{max}=4\pi$ are
    indicated by green lines. The dashed lines indicate thresholds for on-shell
    decays of one heavy 2HDM Higgs boson into other heavy 2HDM Higgs bosons.}
  \label{fig:mh-full}
\end{figure}
The implications of the current experimental data for the masses of the heavy
2HDM Higgs bosons are summarised in Fig.~\ref{fig:mh-full}.
Fig.~\ref{fig:mh-full}(a) shows, as a function of $m_\phi$, the $p$-value for
the hypothesis that a certain heavy 2HDM Higgs boson $\phi$ ($=H,A,H^\pm$) has a
certain mass $m_\phi$. In addition to the lower limits on $m_{H^\pm}$ which were
already shown in Fig.~\ref{fig:tb_bma_mHp-full}(b) we see that masses of the
heavy neutral Higgs bosons below approximately \unit{375}{GeV} are disfavoured
at one standard deviation. At two standard deviations all values down to
\unit{126}{GeV} (in the case of $m_H$) or below (in the case of $m_A$) are
allowed. However, certain \emph{combinations} of heavy Higgs masses can be
excluded with a higher significance. This is shown in Figs.~\ref{fig:mh-full}(b)
to (d). The dashed lines indicate the thresholds for various tree-level
$\phi\to\phi'\phi''$ and $\phi\to\phi'V$ decays (with
$\phi,\phi',\phi''\in\{H,A,H^\pm\}$ and $V\in\{W,Z\}$).
Fig.~\ref{fig:mh-full}(b) shows that scenarios where both $m_H$ and $m_A$ are
smaller than \unit{300}{GeV} are excluded at two standard deviations. In
Fig.~\ref{fig:mh-full}(d) we see that the lower limit of $m_{H^\pm}$ increases
slightly for values of $m_H$ below approximately \unit{400}{GeV}. Both limits
come from the combination of flavour and electroweak precision observables and
are independent of the implementation of the perturbativity bound. For
$m_H<m_{H^\pm}$ the EWPO constraints can only be satisfied if $m_A\approx
m_{H^\pm}$. Combined with the lower bound on $m_{H^\pm}$ from $\br(B\to
X_s\gamma)$ this explains the exclusion of the lower left corner in the
$m_H$-$m_A$ plane. Furthermore, the top-left and bottom-right regions in
Figs.~\ref{fig:mh-full}(b) to (d) are excluded because the requirement of
perturbativity constrains the differences between the heavy Higgs masses to be
of order $v$. Naturally, these limits depend on the implementation of the
perturbativity bound.  For the tight bound ($\lambda_\text{max}=2\pi$) the
(on-shell) decay $H\to H^+H^-$ is excluded at two standard deviations.

%%%%%%%%%%%%%%%%%%%%%%%%%%%%%%%%%%%%%%%%%%%%%%%%%%%%%%%%%%%%%%%%%%%%%%%%%%%%%%%%
\section{Conclusions\label{sect:c}}
%%%%%%%%%%%%%%%%%%%%%%%%%%%%%%%%%%%%%%%%%%%%%%%%%%%%%%%%%%%%%%%%%%%%%%%%%%%%%%%%

In this paper we have confronted the type-II 2HDM (with a softly broken $Z_2$
symmetry) with the relevant experimental constraints from LHC data on the
\unit{126}{GeV} Higgs resonance, the non-observation of additional heavy Higgs
resonances, electroweak precision and flavour observables. In addition
theoretical constraints from the requirements of vacuum stability and
perturbativity were taken into account. While the requirement for perturbativity
of the Higgs self-couplings must be included in some way, we emphasise that the
definition of the perturbativity bound involves some arbitrariness. Therefore,
the approach taken in this paper is to show results for both a loose
($\lambda_\text{max}=4\pi$) and a tight ($\lambda_\text{max}=2\pi$)
implementation of this bound.

In the present analysis the \unit{126}{GeV} resonance is always interpreted as
the light $CP$-even 2HDM Higgs boson. We find that the combination of Higgs
signal strength data and flavour observables disfavours, at one standard
deviation, scenarios where the couplings of light $CP$-even Higgs boson deviate
strongly from the ones of the SM Higgs boson. (We are referring to the $2\sigma$
`islands' in Fig.~\ref{fig:tb_bma_mHp-full}(a) and (c).) For $\tan\beta<5$ such
scenarios are excluded at two standard deviations. Charged Higgs masses below
\unit{322}{GeV} are also excluded at two standard deviations. This limit is
mainly due to the $\br(\bar B\to X_s\gamma)$ measurement and our fit uses the
most accurate available theoretical computation \cite{Hermann:2012fc,
  Misiak:2006ab} of this quantity. Furthermore, flavour and electroweak
precision observables exclude scenarios with both $m_H$ and $m_A$ below
approximately \unit{300}{GeV} at two standard deviations.  For large values of
$m_H$, $m_A$ and $m_{H^\pm}$ the differences between these masses are bounded by
the requirement of perturbativity. If the tight version
($\lambda_\text{max}=2\pi$) of the perturbativity bound is employed, the
on-shell $H\to H^+H^-$ decay is ruled out at two standard deviations.

Our results differ from several recent analyses of the 2HDM of type
II.  Contrary to statements in e.g.~\cite{Grinstein:2013npa, Barroso:2013zxa} we
find that scenarios where $\beta-\alpha$ deviates significantly from $\pi/2$ are
disfavoured at one standard deviation and excluded at two standard deviations
for $\tan\beta<5$. This exclusion is a consequence of the combination of light
Higgs signal strengths with flavour observables.  We also do not confirm the
upper limits on the heavy Higgs masses reported in \cite{Coleppa:2013dya}. As
explained in \cite{Gunion:2002zf} the type-II 2HDM with a softly broken $Z_2$
symmetry has a decoupling limit in which the light $CP$-even Higgs boson becomes
SM-like and the other Higgs bosons become infinitely heavy. In this limit the
theory is phenomenologically indistinguishable from the SM and perturbativity of
the Higgs self-couplings enforces precise relations between the heavy Higgs
masses. The scan-based analysis of \cite{Coleppa:2013dya} simply misses the
scenarios where these relations are fulfilled. For the same reason, we do not
confirm the upper limit on $\tan\beta$ reported there.

\section*{Acknowledgements}
We would like to thank Thomas Hermann, Mikolaj Misiak and Matthias Steinhauser
for sharing their code for $\br(\bar B\to X_s\gamma)$ with us. We also thank
Rui Santos for fruitful discussions and the CKMfitter group for access
to and support for their analysis framework.

\bibliography{thdmfit}
\input{inputs.tex}

\end{document}